
\magnification = \magstep1
\hsize=13cm
\vskip3.5cm
\centerline{\bf Analytic Bethe Ansatz and $T$-system in $C^{(1)}_2$ vertex
models}
\vskip1cm
\noindent
\centerline{by}\par
\vskip1cm
\centerline{{\bf A. KUNIBA}\footnote\dag{E-mail:
kuniba@math.sci.kyushu-u.ac.jp}}
\vskip1cm
\noindent
\centerline{\sl Department of Mathematics,}
\centerline{\sl Faculty of Science, Kyushu University,}
\centerline{\sl Fukuoka 812 Japan}
\vskip0.4cm
\noindent

\vskip0.4cm\noindent
\vskip6cm\noindent
\centerline{\bf ABSTRACT}\par\vskip0.2cm
Eigenvalues of the commuting family of transfer matrices
are expected to obey the $T$-system, a set of functional relation,
proposed recently.
Here we obtain the solution to the $T$-system
for $C^{(1)}_2$ vertex models.
They are compatible with the analytic Bethe ansatz and
Yang-Baxterize the classical characters.

\vfill\eject
\noindent
Solvable lattice models in two-dimensions
possess a commuting family of the row-to-row
transfer matrices [1].
Recently, a set of functional relations (FRs),
the {\it $T$-system},
are proposed among them [2] for a
wide class of models associated with any classical simple Lie algebra or
its quantum affine analogue [3,4].
In the QISM terminology [5],
the $T$-system relates the transfer matrices
with various fusion types in the auxiliary space
but acting on a common quantum space.
It generalizes earlier FRs [6-9]
and enables the calculation of various physical quantities [10].
The structure that underlies the $T$-system is
an (short) exact sequence of the finite dimensional modules of the
above mentioned algebras [2].
As discussed therein, there is an intrigueing connection
between the $T$-system, the thermodynamic Bethe ansatz (TBA) and
dilogarithm identities, indicating some deep interplay among these subjects.
\par
In this Letter we report the solution to the
$C^{(1)}_2$ $T$-system that is compatible with
the analytic Bethe ansatz [11,12] and Yang-Baxterizes
the classical characters.
To explain the problem, let $W^{(a)}_m$
($a = 1, 2, m \in {\bf Z}_{\ge 1}$)
be the irreducible finite dimensional representation (IFDR)
of the quantum affine algebra $U_q(C^{(1)}_2)$ ($q$: generic)
as sketched in section 3.2 of [2].
As the $C_2$-module, it decomposes as
$$\eqalign{
W^{(1)}_m &\simeq V_{m\omega_1} \oplus V_{(m-2)\omega_1} \oplus
\cdots \oplus \cases{V_0& $m$ even\cr V_{\omega_1}& $m$ odd},
\cr
W^{(2)}_m &\simeq V_{m\omega_2},}
\eqno(1)
$$
where $\omega_1, \omega_2$ are the fundamental weights and
$V_\omega$ denotes the IFDR of $C_2$ with highest weight $\omega$.
Thus $\hbox{dim} W^{(1)}_m = (m+2)(m+4)(m^2+6m+6)/48$
for $m$ even, $= (m+1)(m+3)^2(m+5)/48$ for $m$ odd and
$\hbox{dim} W^{(2)}_m = (m+1)(m+2)(2m+3)/6$.
For $W, W^\prime \in \{W^{(a)}_m \mid, a=1,2, m \in {\bf Z}_{\ge 1} \}$,
there exists the quantum $R$-matrix
$R_{W,W^\prime}(u)$ acting on $W\otimes W^\prime$ and satisfying
the Yang-Baxter equation
$$R_{W,W^\prime}(u)R_{W,W^{\prime\prime}}(u+v)R_{W^\prime,W^{\prime\prime}}(v)
= R_{W^\prime,W^{\prime\prime}}(v)R_{W,W^{\prime\prime}}(u+v)
R_{W,W^\prime}(u)\eqno(2)$$
with $u, v \in {\bf C}$ being the spectral parameters.
For $W = W^\prime = W^{(1)}_1$, the $R$-matrix has been explicitly
written down in [13,14], from which all the other $R_{W, W^\prime}$ may be
constructed via the fusion procedure [15].
$W^{(a)}_m$ is an analogue of the $m$-fold symmetric tensor
representation of $W^{(a)}_1$.
The transfer matrix with auxiliary space $W^{(a)}_m$ is then
defined by
$$T^{(a)}_m(u) = \hbox{Tr}_{W^{(a)}_m}\Bigl(
R_{W^{(a)}_m,W^{(p)}_s}(u-w_1) \cdots
R_{W^{(a)}_m,W^{(p)}_s}(u-w_N)\Bigr) \eqno(3)$$
up to an overall scalar multiple.
Here $N \in 2{\bf Z}$ denotes the system size, $w_1, \ldots w_N$ are
complex parameters representing the inhomogeneity,
$p=1,2$ and $s \in {\bf Z}_{\ge 1}$.
We say that (3) is the row-to-row transfer matrix with fusion type
$W^{(a)}_m$ acting on the quantum space $(W^{(p)}_s)^{\otimes N}$.
We shall reserve the letters $p$ and $s$ for this meaning
throughout.
Thanks to the Yang-Baxter equation (2), the transfer matrices (3) form a
commuting family
$$[T^{(a)}_m(u), T^{(a^\prime)}_{m^\prime}(u^\prime)] = 0.\eqno(4)$$
We shall write the eigenvalues of $T^{(a)}_m(u)$ as
$\Lambda^{(a)}_m(u)$.
Our goal is to find an exlicit formula for them.
\par
For the purpose, we postulate the (unrestricted) $T$-system [2]:
$$\eqalignno{
T^{(1)}_{2m}(u-{1\over 2})T^{(1)}_{2m}(u+{1\over 2})
&= T^{(1)}_{2m+1}(u)T^{(1)}_{2m-1}(u) \cr
&+ g^{(1)}_{2m}(u)
T^{(2)}_m(u-{1\over 2})T^{(2)}_m(u+{1\over 2}),&(5{\rm a})\cr
T^{(1)}_{2m+1}(u-{1\over 2})T^{(1)}_{2m+1}(u+{1\over 2})
&= T^{(1)}_{2m+2}(u)T^{(1)}_{2m}(u) \cr
&+ g^{(1)}_{2m+1}(u)
T^{(2)}_m(u)T^{(2)}_{m+1}(u),&(5{\rm b})\cr
T^{(2)}_m(u-1)T^{(2)}_m(u+1) &=
T^{(2)}_{m+1}(u)T^{(2)}_{m-1}(u)
+ g^{(2)}_{m}(u)T^{(1)}_{2m}(u).&(5{\rm c})\cr}
$$
Here $g^{(a)}_m(u)$ is a scalar function that depends on
$W^{(p)}_s$ and overall normalization of the transfer matrices.
Due to (4) the eigenvalues $\Lambda^{(a)}_m(u)$ also obey the
same system as (5), which can be solved succesively yielding an expression
of the $\Lambda^{(a)}_m(u)$ in terms of
$\Lambda^{(1)}_1(u+\hbox{shift})$ and $\Lambda^{(2)}_1(u+\hbox{shift})$.
Thus the first step to achieve the goal is to find the
formula for the eigenvalues $\Lambda^{(1)}_1(u)$ and $\Lambda^{(2)}_1(u)$.
This we do by the analytic Bethe ansatz.
The method consists of assuming the so-called ``dressed vacuum form"
for the eigenvalues and determining the unknown parts thereby introduced
from some functional properties and asymptotic behaviors.
See [12,16] for the detail.
To present the results for our problem, we prepare a few notations.
Let $\alpha_1, \alpha_2$ be the simple roots of $C_2$.
We take $\alpha_2$ to be a long root and normalize it as
$(\alpha_2\vert \alpha_2) = 2$ via the bilinear form $(\, \vert \, )$.
Then one has
$(\alpha_a \vert \omega_b) = \delta_{a b}/t_a$, where
$t_1 = 2,\, t_2 = 1$.
We set
$$\eqalign{
\phi(u) &= \prod_{j=1}^N [u - w_j], \quad [u] = q^u - q^{-u},\cr
\phi^{(a)}_m(u) &= \phi(u+{m-1\over t_a})\phi(u+{m-3\over t_a}) \cdots
\phi(u-{m-1\over t_a})\quad a = 1, 2,\, m \in {\bf Z}_{\ge 1},\cr
Q_a(u) &= \prod_{j=1}^{N_a}[u - iu^{(a)}_j]\quad a = 1, 2.\cr
}\eqno(6)$$
Here $N_1$, $N_2$ are non-negative integers such that
$\omega^{(p)}
\buildrel\rm def\over = Ns\omega_p - N_1\alpha_1 - N_2\alpha_2$ is a
non-negative weight.
The numbers $\{u^{(a)}_j \mid a=1, 2, 1 \le j \le N_a \}$
are the solutions to the Bethe ansatz equation [16]
$$\eqalign{
-{\phi(iu^{(a)}_k+{s\over t_p}\delta_{p a})\over
\phi(iu^{(a)}_k-{s\over t_p}\delta_{p a})} &=
\prod_{b=1}^2{Q_b(iu^{(a)}_k+(\alpha_a \vert \alpha_b))\over
Q_b(iu^{(a)}_k-(\alpha_a \vert \alpha_b))},
\quad a = 1, 2, \, 1 \le k \le N_a.\cr}
\eqno(7)
$$
Under these definitions, the result of the analytic Bethe ansatz reads
as follows.
$$\eqalignno{\hbox{Case } p &= 1;\cr
\Lambda^{(1)}_1(u) &=
\phi^{(1)}_s(u+3)\phi^{(1)}_s(u+1){Q_1(u-{1\over 2})\over Q_1(u+{1\over 2})}
+ \phi^{(1)}_s(u+2)\phi^{(1)}_s(u){Q_1(u+{7\over 2})\over Q_1(u+{5\over 2})}\cr
&+ \phi^{(1)}_s(u+3)\phi^{(1)}_s(u)\Bigl({Q_1(u+{3\over 2})Q_2(u-{1\over
2})\over
Q_1(u+{1\over 2})Q_2(u+{3\over 2})}
+{Q_1(u+{3\over 2})Q_2(u+{7\over 2})\over
Q_1(u+{5\over 2})Q_2(u+{3\over 2})}
\Bigr), &(8{\rm a})\cr
\Lambda^{(2)}_1(u) &=
\phi^{(1)}_s(u+{5\over 2})\Bigl(
{Q_2(u-1)\over Q_2(u+1)} + {Q_1(u)Q_2(u+3)\over Q_1(u+2)Q_2(u+1)}\Bigr)\cr
&+ \phi^{(1)}_s(u+{1\over 2})\Bigl(
{Q_2(u+4)\over Q_2(u+2)} + {Q_1(u+3)Q_2(u)\over Q_1(u+1)Q_2(u+2)}\Bigr)\cr
&+ \phi^{(1)}_s(u+{3\over 2})
{Q_1(u)Q_1(u+3)\over Q_1(u+1)Q_1(u+2)}.&(8{\rm b})\cr
\hbox{Case } p &= 2;\cr
\Lambda^{(1)}_1(u) &= \phi^{(2)}_s(u+{5\over 2})\phi^{(2)}_s(u+{3\over 2})
\Bigl({Q_1(u-{1\over 2})\over Q_1(u+{1\over 2})} +
{Q_1(u+{3\over 2})Q_2(u-{1\over 2})\over
Q_1(u+{1\over 2})Q_2(u+{3\over 2})}\Bigr)\cr
&+ \phi^{(2)}_s(u+{3\over 2})\phi^{(2)}_s(u+{1\over 2})
\Bigl({Q_1(u+{3\over 2})Q_2(u+{7\over 2})\over
Q_1(u+{5\over 2})Q_2(u+{3\over 2})} +
{Q_1(u+{7\over 2})\over Q_1(u+{5\over 2})}\Bigr),&(8{\rm c})\cr
\Lambda^{(2)}_1(u) &=
\phi^{(2)}_s(u+3)\phi^{(2)}_s(u+2){Q_2(u-1)\over Q_2(u+1)} +
\phi^{(2)}_s(u+1)\phi^{(2)}_s(u){Q_2(u+4)\over Q_2(u+2)}\cr
&+ \phi^{(2)}_s(u+3)\phi^{(2)}_s(u)\Bigl(
{Q_1(u)Q_2(u+3)\over Q_1(u+2)Q_2(u+1)} + {Q_1(u)Q_1(u+3)\over
Q_1(u+1)Q_1(u+2)}\cr
&\qquad\qquad\qquad\qquad\quad +
{Q_1(u+3)Q_2(u)\over Q_1(u+1)Q_2(u+2)}\Bigr).&(8{\rm d})\cr}
$$
We employ the convention such that
the eigenvalue of $\check{R}_{W^{(1)}_1,W^{(1)}_1}(u)$
on the highest component $V_{2\Lambda_1}$ is
$[u+3][u+1]$ and let the overall normalization
of $\Lambda^{(a)}_1(u)$ as specified by (8).
(The common factor $\phi^{(2)}_s(u+{3\over 2})$ in (8c)
has been attached so as to simplify the forthcoming formula (12).)
The $\Lambda^{(a)}_1(u)$ consists of
$\hbox{dim} W^{(a)}_1 = 4, 5 \,(a = 1, 2)$ terms and
its pole-free conditions are given by (7)
in accordance with the analytic Bethe ansatz.
The formulas (8) coincide with those
in [16,17] for some special cases.
In particular, ratio of $Q_a$'s in $\Lambda^{(1)}_1(u)$
are just those appearing in [16] for the
$C^{(1)}_2$ vertex model with $W^{(p)}_s = W^{(1)}_1$
(upon some convention adjustment).
Namely, the $Q_a$-part is determined only from
the auxiliary space choice, while the quantum space
dependence enters $\phi^{(p)}_s$-part.
This is also the case in the formula
eq.(3.17) of [8] for the $sl(n)$ case.
Similary, $Q_a$-part in $\Lambda^{(2)}_1(u)$ are
those appearing in the $B^{(1)}_2$ case of [16]
due to the equivalence $C_2 \simeq B_2$.
\par
To proceed to $\Lambda^{(a)}_m(u)$ with higher $m$,
we introduce a few more notations.
$$\eqalign{
G_a(u) &= \cases{\phi^{(1)}_s(u)G(u)& for $a = 1$\cr
                    G(u)& for $a = 2$\cr},\quad
H_a(u) = \cases{H(u)& for $a = 1$\cr
                    \phi^{(2)}_s(u)H(u)& for $a = 2$\cr},\quad\cr
G(u) &= {Q_2(u+{1\over 2})Q_2(u-{1\over 2})\over
Q_1(u+{1\over 2})Q_1(u-{1\over 2})},\quad
H(u) = {Q_1(u)\over Q_2(u+1)Q_2(u-1)}.\cr}\eqno(9)
$$
We consider the $T$-system (5)
for $\Lambda^{(a)}_m(u)$ with the initial condition for $m=1$ as (8) and
$$\eqalign{
\Lambda^{(1)}_0(u) &= \phi^{(1)}_s(u+5/2)\phi^{(1)}_s(u+1/2),\,
\Lambda^{(2)}_0(u) = \phi^{(1)}_s(u+3/2)\,\, \hbox{ for }\,\, p = 1,\cr
\Lambda^{(1)}_0(u) &= \Lambda^{(2)}_0(u) = \phi^{(2)}_s(u+1)\phi^{(2)}_s(u+2)
\,\, \hbox{ for }\,\, p = 2.\cr}\eqno(10)
$$
Then our main result is
\proclaim Theorem.
The functions
$$\eqalign{
\Lambda^{(1)}_m(u) &= Q_1(u-{m\over 2})Q_1(u+{m\over 2}+3)
\sum_{0 \le i \le j \le m}
\sum_{l=[{i+1\over 2}]}^{[{j+1\over 2}]}
\sum_{k=[{i\over 2}]}^{[{j\over 2}]}
G_p(u+{m+5\over 2}-i)\cr
&\qquad\qquad\qquad G_p(u+{m+1\over 2}-j)
H_p(u+{m\over 2}-2l+2)H_p(u+{m\over 2}-2k+1),\cr
\Lambda^{(2)}_m(u) &= Q_2(u-m)Q_2(u+m+3)
\sum_{j=0}^{2m}\sum_{l=0}^{[{j\over 2}]}
\sum_{k=[{j+1\over 2}]}^m
G_p(u+m+{3\over 2}-j)\cr
&\qquad\qquad\qquad H_p(u+m-2l+2)H_p(u+m-2k+1)}\eqno(11)
$$
are the solutions to the
$T$-system (5) with the intial condition (8), (10)
and $g^{(a)}_m(u)$ given by
$$g^{(a)}_m(u)
= \cases{\phi^{(p)}_s(u+{m\over t_p}+3)\phi^{(p)}_s(u-{m\over t_p})& if $a =
p$\cr
         1& otherwise\cr}.\eqno(12)
$$

The symbol $[x]$ in (11) denotes the greatest integer not exceeding $x$
and should not be confused with the one in (6).
The function (12) satisfies
$g^{(a)}_m(u-{1\over t_a})g^{(a)}_m(u+{1\over t_a})
= g^{(a)}_{m+1}(u)g^{(a)}_{m-1}(u)$ in accordance with
eq.(3.18) of [2] (with a slight normalization change in $u$).
The theorem can be proved by comparing the
coefficients of $\phi^{(a)}_s$ factors on both sides of the
$T$-system.
In particular, the check essentially reduces to the case
$p = s = 1$.
A similar formula to (11) is available for
$sl(n)$ case in [8].
\par
$\Lambda^{(a)}_m(u)$ (11) Yang-Baxterizes
the character of
$W^{(a)}_m$ viewed as a $C_2$-module as in (1).
Namely, it contains
$\hbox{ dim } W^{(a)}_m$ terms and tends to the latter
in the ``braid limit" as follows.
$$\eqalign{
&\lim_{u \rightarrow \infty, (\vert q \vert > 1)}
q^{-\psi_a} \Lambda^{(a)}_m(u)
= \chi^{(a)}_m(q^{(\omega^{(p)},\alpha_1)},q^{(\omega^{(p)},\alpha_2)}),\cr
&\psi_a = s(2Nu+3N-2\sum_{j=1}^N w_j)\hbox{min }(1, {t_a\over t_p}),\cr
&\chi^{(1)}_m(z_1,z_2) = \sum_{0 \le i \le j \le m}
\sum_{l=[{i+1\over 2}]}^{[{j+1\over 2}]}
\sum_{k=[{i\over 2}]}^{[{j\over 2}]}
z_1^{2m-2i-2j}z_2^{m-2l-2k}\cr
&\qquad = \hbox{ch }V_{m\omega_1} + \hbox{ch }V_{(m-2)\omega_1}
+ \cdots + \cases{1 & $m$ even\cr \hbox{ch }V_{\omega_1} & $m$ odd\cr},\cr
&\chi^{(2)}_m(z_1,z_2) =
\sum_{j=0}^{2m}\sum_{l=0}^{[{j\over 2}]}
\sum_{k=[{j+1\over 2}]}^m
z_1^{2m-2j}z_2^{2m-2l-2k}\cr
&\qquad = \hbox{ch }V_{m\omega_2}.\cr}\eqno(13)
$$
Here, $\hbox{ch }V_\omega = \hbox{ch }V_\omega(z_1,z_2)$ is the irreducible
$C_2$ character with highest weight $\omega$
counting the $(\xi\alpha_1 + \eta\alpha_2)$-weight vectors
as $z_1^{2\xi}z_2^{2\eta}$.
The following character identity [18]
is a simple corollary of the above theorem.
$$\eqalign{
\chi^{(1) 2}_{2m}
&= \chi^{(1)}_{2m+1}\chi^{(1)}_{2m-1} + \chi^{(2) 2}_m,\cr
\chi^{(1) 2}_{2m+1}
&= \chi^{(1)}_{2m+2}\chi^{(1)}_{2m} +
\chi^{(2)}_m\chi^{(2)}_{m+1},\cr
\chi^{(2) 2}_m &=
\chi^{(2)}_{m+1}\chi^{(2)}_{m-1}
+ \chi^{(1)}_{2m}.\cr}\eqno(14)
$$
In [2,10], $\chi^{(a)}_m$ was denoted by $Q^{(a)}_m$ and (14) was called
the $Q$-system.
As shown therein, the combinations
$y^{(1)}_{2m}(u)
= {g^{(1)}_{2m}(u)\Lambda^{(2)}_m(u-{1\over2})\Lambda^{(2)}_m(u+{1\over 2})
\over \Lambda^{(1)}_{2m+1}(u)\Lambda^{(1)}_{2m-1}(u)}$ etc from (5) yield
a solution to the $C^{(1)}_2$ $Y$-system [19], the TBA equation in high
temperature
limit:
$$\eqalign{
&y^{(1)}_{2m}(u+{1\over 2})y^{(1)}_{2m}(u-{1\over 2})
= {1 + y^{(2)}_m(u) \over
(1 + y^{(1)}_{2m-1}(u)^{-1})(1 + y^{(1)}_{2m+1}(u)^{-1})},\cr
&y^{(1)}_{2m+1}(u+{1\over 2})y^{(1)}_{2m+1}(u-{1\over 2})
= {1 \over
(1 + y^{(1)}_{2m+2}(u)^{-1})(1 + y^{(1)}_{2m}(u)^{-1})},\cr
&y^{(2)}_m(u+1)y^{(2)}_m(u-1)\cr
&\qquad\qquad= {(1 + y^{(1)}_{2m-1}(u))
(1 + y^{(1)}_{2m}(u+{1\over 2}))(1 + y^{(1)}_{2m}(u-{1\over 2}))
(1 + y^{(1)}_{2m+1}(u)) \over
(1 + y^{(2)}_{m-1}(u)^{-1})(1 + y^{(2)}_{m+1}(u)^{-1})}.\cr
}\eqno(15)
$$
\vskip0.5cm
\beginsection Acknowledgement

The author thanks Junji Suzuki for a useful discussion and critical
reading of the manuscript.
\vfill\eject
\beginsection Reference

\item{[1]}{R.J.Baxter, {\it Exactly solved models in statistical mechanics},
(Academic Press, London, 1982)}
\item{[2]}{A.Kuniba, T.Nakanishi and J.Suzuki, ``Functional relations
in solvable lattice models I: Functional relations and representation theory",
HUTP-93/A022, hep-th.9309137}
\item{[3]}{V.G.Drinfel'd, in {\it Proceedings of the ICM, Berkeley}
(AMS, Providence, 1987)}
\item{[4]}{M.Jimbo, Lett. Math. Phys. {\bf 10} (1985) 63}
\item{[5]}{P.P.Kulish and E.K.Sklyanin, in
{\it Lecture Notes in Physics} {\bf 151} (Springer, Berlin 1982)}
\item{[6]}{R.J.Baxter and
P.A.Pearce, J.Phys. A: Math. Gen.  {\bf 15} (1982) 897}
\item{[7]}{V.V.Bazhanov and N.Yu.Reshetikhin,
Int. J. Mod. Phys. {\bf A4} (1989) 115}
\item{[8]}{V.V.Bazhanov and N.Yu.Reshetikhin,
J. Phys. A: Math. Gen. {\bf 23} (1990) 1477}
\item{[9]}{A.Kl\"umper and P.A.Pearce, Physica {\bf A183} (1992) 304}
\item{[10]}{A.Kuniba, T.Nakanishi and J.Suzuki, ``Functional relations
in solvable lattice models II: Applications",
HUTP-93/A023, hep-th.9310060}
\item{[11]}{R.J.Baxter, Ann. Phys. {\bf 70} (1972) 193}
\item{[12]}{N.Yu.Reshetikhin, Sov. Phys. JETP {\bf 57} (1983) 691}
\item{[13]}{V.V.Bazhanov, Phys. Lett. {\bf B159} (1985) 321}
\item{[14]}{M.Jimbo, Commun. Math. Phys. {\bf 102} (1986) 537}
\item{[15]}{P.P.Kulish, N.Yu.Reshetikhin and E.K.Sklyanin,
Lett. Math. Phys. {\bf 5} (19981) 393}
\item{[16]}{N.Yu.Reshetikhin, Lett. Math. Phys. {\bf 14} (1987) 235}
\item{[17]}{N.Yu.Reshetikhin, Theor. Math. Phys. {\bf 63} (1985) 347}
\item{[18]}{A.N.Kirillov and N.Yu.Reshetikhin, J. Sov. Math. {\bf 52}
(1990) 3156}
\item{[19]}{A.Kuniba and T.Nakanishi, Mod. Phys. Lett. {\bf A7} (1992) 3487}
\bye